\documentclass[copyright,creativecommons]{eptcs}
\providecommand{\event}{ICE 2010} % Name of the event you are submitting to
\usepackage{breakurl}             % Not needed if you use pdflatex only.

\usepackage{amsmath}
\usepackage{amssymb}
\usepackage{units}
\usepackage{wasysym}
\usepackage{graphics}
\usepackage{color}

\newcommand{\comment}[1]{}

\newcommand{\pcl}{\textup{PCL\;}}
\newcommand{\coimp}{\twoheadrightarrow}
\newcommand{\imp}{\rightarrow}
\newcommand{\nrule}[1]{{\footnotesize \textsc{#1}}}
\newcommand{\irule}[2]{\frac{\textstyle\rule[-1.3ex]{0cm}{3ex}#1}%
{\textstyle\rule[-.5ex]{0cm}{3ex}#2}}
\newcommand{\powset}[1]{\mathcal{P}(#1)}

\newcommand{\bnfdef}{::=}
\newcommand{\bnfmid}{\;\big|\;}

\newcommand{\ask}[2][]{\mathsf{ask}_{#1}\,{#2}}
\newcommand{\join}[2]{\mathsf{join}_{#1}\,{#2}}
\newcommand{\fuse}[2]{\mathsf{fuse}_{#1}\,{#2}}
\newcommand{\checkp}[1]{\mathsf{check}\,#1}

\global\long\def\tell#1{\mathsf{tell}\,#1}
\global\long\def\checkp#1{\mathsf{check}\,#1}

\newcommand{\rew}{\rightarrow}

\newcommand{\bind}[2]{\nicefrac{#2}{#1}}
\newcommand{\setenum}[1]{\{#1\}}

\newtheorem{theorem}{Theorem}

\newtheorem{example}{Example}
\newtheorem{definition}{Definition}

\def\titlerunning{Primitives for Contract-based Synchronization}
\def\authorrunning{M. Bartoletti \& R. Zunino}

\begin{document}

\title{Primitives for Contract-based Synchronization}

\author{Massimo Bartoletti
\institute{Dipartimento di Matematica e Informatica, Universit\`a degli Studi di Cagliari}
% \email{bart@unica.it}
\and
Roberto Zunino
\institute{Dipartimento di Ingegneria e Scienza dell'Informazione, Universit\`a degli Studi di Trento}
% \email{zunino@disi.unitn.it}
}

\maketitle

\begin{abstract}
We investigate how contracts can be used to regulate the
interaction between processes.
To do that, we study a variant of the concurrent constraints calculus
presented in~\cite{BZ10lics} , featuring primitives for
multi-party synchronization via contracts.
We proceed in two directions.
First, we exploit our primitives to model some contract-based interactions.
Then, we discuss how several models for concurrency can be expressed 
through our primitives. 
In particular, we encode the \mbox{$\pi$-calculus} and graph rewriting.
\end{abstract}

\newcommand{\airplane}[0]{a}
\newcommand{\bike}[0]{b}

\section{Introduction} \label{sec:intro}

A contract is a binding agreement stipulated between two or more parties, 
which dictates their rights and their duties, 
and the penalties each party has to pay in case the contract is not honoured.

In the current practice of information technology, contracts are not that
different from those legal agreements traditionally enforced in courts of law.
Both software and services commit themselves to respect some 
(typically weak, if not ``without any expressed or implied warranty'') 
service level agreement. 
In the case this is not honoured, the
only thing the user can do is to take legal steps against 
the software vendor or service provider. 
Since legal disputes may require a lot of time, as well as relevant expenses, 
such kinds of contracts serve more as an instrument to discourage users,
rather than making easier for users to demand their rights.

Recent research has then addressed the problem of devising new kinds of
contracts, to be exploited for specifying and automatically regulating 
the interaction among users and service providers.
See e.g.~\cite{Bruni09models,Buscemi07ccpi,Carpineti06basic,Castagna09mobile,Padovani09contract}, 
to cite a few.
A contract subordinates the behaviour promised by a client 
(e.g.\ ``I will pay for a service~X'')
to the behaviour promised by a service
(e.g.\ ``I will provide you with a service~Y''),
and \emph{vice versa}.
The crucial problems are then how to formalise the concept of contract, 
how to understand when a set of contracts gives rise to an agreement
among the stipulating parties,
and how to actually enforce this agreement in an open, and possibly unreliable,
environment.

In the Concurrent Constraint Programming (CCP)
paradigm \cite{Saraswat93cc,Saraswat91cc},
concurrent processes communicate through a global constraint store.
A process can add a constraint $c$ to the store through 
the $\tell{c}$ primitive.
Dually, the primitive $\ask{c}$ makes a process block until
the constraint $c$ is entailed by the store.
Very roughly, such primitives may be used to model two basic operations on
contracts: a $\tell{c}$ is for publishing the contract $c$, and
an $\ask{c'}$ is for waiting until one has to fulfill some duty $c'$.

While this may suggest CCP
as a good candidate for modelling contract-based interactions,
some important features seem to be missing.
Consider e.g.\ a set of parties, each offering her own contract.
When some of the contracts at hand give rise to an agreement,
all the involved parties accept the contract, and start interacting 
to accomplish it.
A third party (possibly, an ``electronic'' court of law) may later on
join these parties, so to provide the stipulated remedies in the case 
an infringement to the contract is found.
To model this typical contract-based dynamics, we need
the ability of making \emph{all} the parties involved
in a contract synchronise when an agreement is found,
establishing a session.
Also, we need to allow an external party to join a running session, according
to some condition on the status of the~contract.

In this paper we study a variant of~\cite{BZ10lics}, an extension of CCP 
which allows for modelling such kinds of scenarios.
Our calculus features two primitives, called $\fuse{}{}$ and $\join{}{}\!$:
the first fuses all the processes agreeing on a given contract, while
the second joins a process with those already participating to a contract.
Technically, the prefix $\fuse{x}{c}$ probes the constraint store 
to find whether it entails~$c$; 
when this happens, the variable~$x$ is bound 
to a fresh session identifier, 
shared among the parties involved in the contract.
Such parties are chosen according to a {\em local minimal fusion} policy.
The prefix $\join{x}{c}$ is similar, yet it looks for an already existing
session identifier, rather than creating a fresh one.
While our calculus is undogmatic about the underlying constraint system,
in the contract-based scenarios presented here we commit ourselves 
to using \pcl formulae~\cite{BZ10lics} as constraints.

\paragraph{Contributions.}
Our contribution consists of the following points.
In Sect.~\ref{sec:calculus} we study a calculus for contracting processes.
Compared to the calculus in~\cite{BZ10lics},
the one in this paper
differs in the treatment of the main primitives $\fuse{}{}$ and $\join{}{}$\!,
which have a simplified semantics.
Moreover, we also provide here a reduction semantics, 
and compare it to the labelled one.
In Sect.~\ref{sec:examples} we show our calculus suitable for modelling 
complex interactions of contracting parties. 
In Sect.~\ref{sec:encodings} we substantiate a statement made 
in~\cite{BZ10lics}, by showing how to actually encode into our calculus
some common concurrency idioms, 
among which the $\pi$-calculus~\cite{Milner92pi} 
and graph rewriting~\cite{Rozenberg97handbook}.
In Sect.~\ref{sec:related-work} we discuss further differences 
between the two calculi, and compare them with other frameworks.

\vspace{-5pt}

\section{A contract calculus}\label{sec:calculus}

We now define our calculus of contracting processes.
The calculus is similar to that in~\cite{BZ10lics},
yet it diverges in the treatment of
the crucial primitives $\fuse{}{}$ and $\join{}{}$.
We will detail the differences between the two versions in
Sect.~\ref{sec:related-work}.
Our calculus features both {\em names}, ranged over by $n,m,\,\ldots\,$,
and {\em variables}, ranged over by $x,y,\,\ldots\,$.
Constraints are {\em terms} over variables and names, and include 
a special element~$\bot$; the set of constraints $D$ is ranged 
over by $c,d$. Our calculus is parametric with respect to an
arbitrary constraint system $(D,\vdash)$ (Def.~\ref{def:constraint-system}).

\begin{definition}[Constraint system~\cite{Saraswat91cc}] \label{def:constraint-system}
A constraint system is a pair $(D,\vdash)$, 
where $D$ is a countable set,
and $\;\vdash \,\subseteq\, \powset{D} \times D$ is a relation
satisfying:
(i) $C \vdash c$ whenever $c \in C$;
(ii) $C \vdash c$ whenever for all $c' \in C'$ we have
$C \vdash c'$, and $C' \vdash c$;
(iii) for any $c$, $C \vdash c$ whenever $C \vdash \bot$.
\end{definition}

\vspace{-15pt}

\paragraph{Syntax.}

Names in our calculus behave similarly to the names in the $\pi$-calculus: 
that is, distinct names represent distinct concrete objects.
Instead, variables behave as the names in the fusion calculus: 
that is, distinct variables can be bound to the same concrete object, 
so they can be fused.
A \emph{fusion} $\sigma$ is a substitution that maps a set of variables
to a single name.
We write $\sigma = \setenum{\bind{\vec{x}}{n}}$ for the fusion
that replaces each variable in~$\vec{x}$ with the name~$n$.
We use metavariables $a,b,\ldots$ to range over both names and variables.

\begin{definition}[Processes]
The set of prefixes and processes are defined as follows:
\begin{align*}
\pi & \bnfdef \tau \bnfmid \tell c \bnfmid \checkp{c} \bnfmid \ask{c} \bnfmid \join{x}{c} \bnfmid \fuse{x}{c}
&& \textit{(prefixes)} \\
P & \bnfdef c \bnfmid \textstyle \sum_{i\in I} \pi_i.P_i \mid P|P \bnfmid (a)P \bnfmid X(\vec{a})
&& \textit{(processes)}
\end{align*}
\end{definition}

Prefixes $\pi$ include $\tau$ (the silent operation as in CCS),
as well as $\tell{}$, $\checkp{}$ and $\ask{}\!$ 
as in Concurrent Constraints~\cite{Saraswat91cc}. 
The prefix $\tell c$ augments the context with the constraint~$c$. 
The prefix $\checkp c$ checks if~$c$ is consistent with the context. 
The prefix $\ask{c}$ causes a process to stop
until the constraint~$c$ is entailed by the context. 
% Note that, since we only allow negation-free constrainte~$u$ here, 
% the context will always be consistent, by Lemma~\ref{lem:pcl-consistent}. 
The prefixes $\fuse{x}{c}$ and $\join{x}{c}$ 
drive the fusion of the variable~$x$, in two different flavours.  
The prefix $\join{x}{c}$ instantiates $x$ to \emph{any known name}, 
provided that after the instantiation the constraint $c$ is entailed.  
The prefix $\fuse{x}{c}$ fuses $x$ with \emph{any set of known variables},
provided that, when all the fused variables are instantiated to a
fresh name, the constraint $c$ is entailed. 
To avoid unnecessary fusion, the set of variables is required to be minimal
(see Def.~\ref{def:local-minimal-fusion}).
To grasp the intuition behind the two kinds of fusions, 
think of names as session identifiers. 
Then, a $\fuse{x}{c}$ initiates a new session, while 
a $\join{x}{c}$ joins an already initiated session.

Processes $P$ include the constraint $c$, 
the summation $\sum_{i \in I} \pi_i.P_i$ of guarded processes 
over indexing set $I$,
the parallel composition $P|Q$, 
the scope delimitation $(a)P$, 
and the instantiated constant $X(\vec{a})$, 
where $\vec{a}$ is a tuple of names/variables.
When a constraint $c$ is at the top-level of a process,
we say it is \emph{active}.
We use a set of defining equations $\{X_i(\vec{x}) \doteq P_i\}_i$
with the provision that each occurrence of $X_j$ in $P_k$ is guarded,
i.e.~it is behind some prefix.
We shall often use $C=\{c_1,c_2,\ldots\}$ as a process, standing for
$c_1 | c_2 | \cdots $. We write $0$ for the empty sum.
Singleton sums are simply written $\pi.P$.
We use $+$ to merge sums as follows:
\(
\sum_{i\in I} \pi_i.P_i  + 
\sum_{i\in J} \pi_i.P_i  = 
\sum_{i\in I \,\cup\, J} \pi_i.P_i 
\;\;\mbox{if } I\cap J =\emptyset
\).
We stipulate that~$+$ binds more tightly \nolinebreak than $|$. 

Free variables and names of processes are defined as usual: they are
free whenever they occur in a process not under a delimitation. Alpha
conversion and substitutions are defined accordingly. 
As a special case, we let
\(
   (\fuse{x}{c}) \{\nicefrac{n}{x}\} = (\join{x}{c}) \{\nicefrac{n}{x}\} = \ask c \{\nicefrac{n}{x}\}
\).
That is, when a variable $x$ is instantiated to a name, the prefixes
$\fuse{x}{c}$ and $\join{x}{c}$ can no longer require the fusion of $x$, so
they behave as a plain $\ask c$. 
Henceforth, we will consider processes up-to alpha-conversion.

We provide our calculus with both a reduction semantics 
and a labelled transition semantics.
As usual for CCP, the former explains how a process evolves within
the {\em whole} context (so, it is not compositional), 
while the latter also explains how a process interacts
with the environment.

\paragraph{Reduction semantics.}
%The reduction semantics is given in two steps: 
%first, we define a structural equivalence relation between processes.
% (Def.~\ref{def:struct-equiv}),
%Then, we define the reduction relation (Def.~\ref{def:reduction}).
The structural equivalence relation $\equiv$
is the smallest equivalence between processes satisfying the following axioms:
% \begin{figure}
\[\hspace{-5pt}
\begin{array}{@{}c@{}}
P|0\equiv P \quad P|Q\equiv Q|P \quad P|(Q|R)\equiv(P|Q)|R \quad
P+0\equiv P \quad P+Q\equiv Q+P \quad P+(Q+R)\equiv(P+Q)+R \\[10pt]
(a)(P|Q) \equiv P|(a)Q \;\; \mbox{ if }a\not\in \mathit{free}(P) \qquad
(a)(b)P\equiv(b)(a)Q \qquad
X(\vec{a}) \equiv P\{\nicefrac{\vec{a}}{\vec{x}}\} \;\; \mbox{ if $X(\vec{x}) \doteq P$}
\end{array} \\
\]
% \caption{The axioms of structural equivalence.\label{fig:struct-equiv}}
% \end{figure}

% \begin{definition}[Structural equivalence] \label{def:struct-equiv}
%The structural equivalence relation $\equiv$
%is the smallest equivalence between processes satisfying the axioms of
%Fig.~\ref{fig:struct-equiv}.
% \end{definition}

% \[
% \begin{array}{l}
% P | 0 \equiv P \\
% P | Q \equiv Q | P \\
% P | (Q | R) \equiv (P | Q) | R \\
% (a) (P | Q) \equiv (a)P | Q \mbox{ if } a \not\in free(Q) \\
% (a)(b)P \equiv (b)(a)P \\
% X(\vec{a}) \equiv P\{\nicefrac{\vec{a}}{\vec{x}}\} 
% \mbox{ if $X(\vec{x}) \doteq P$}
% \end{array}
% \]
% Note: no choice/delimitation laws since we only have guarded choice.

\begin{definition}[Reduction] \label{def:reduction}
Reduction $\rightarrow$ 
is the smallest relation 
% between processes 
satisfying the rules in Fig.~\ref{fig:reduction}.
\end{definition}

\begin{figure}
\small
\vspace{-10pt}
\[
\begin{array}{c}
  \irule{}{
    (\vec{a}) \, ( \tau.P + Q \mid R ) \rightarrow (\vec{a}) \, ( P \mid R )
}
\;\; \nrule{[Tau]}
\qquad
\irule{}{
(\vec{a}) \, ( \tell c.P + Q \mid R ) \rightarrow (\vec{a}) ( c \mid P \mid R )
}
\;\; \nrule{[Tell]}
\\[20pt]
\irule{
C \vdash c
}{
(\vec{a}) \, (C \mid \ask c . P + Q \mid R)
\rightarrow
(\vec{a}) (C \mid P \mid R)
}
\;\; \nrule{[Ask]}
\qquad
\irule{
C,c \not\vdash \bot \quad R \mbox{ free from active constraints}
}{
(\vec{a}) \, ( C \mid \checkp c.P + Q \mid R ) \rightarrow (\vec{a}) \, ( P \mid R )
}
\;\; \nrule{[Check]}
\\[20pt]
\irule{
\sigma = \setenum{\bind{x\vec{y}}{n}}
\quad
\mbox{$n$ fresh in $P,Q,R,C,c,\vec{a}$}
\quad
C \vdash^{min}_{\sigma} c
}{
(x\vec{y}\vec{a}) \, (C \mid \fuse{x}{c} . P + Q \mid R)
\rightarrow
(n\vec{a}) \, ((C \mid P \mid R) \, \sigma)
} 
\;\; \nrule{[Fuse]}
\\[20pt]
\irule{
C \{\nicefrac{n}{x}\} \vdash c \{\nicefrac{n}{x}\}
}{
(xn\vec{a}) \, (C \mid \join{x}{c} . P + Q \mid R)
\rightarrow
(n\vec{a}) \, \big( (C \mid P \mid R)\{\nicefrac{n}{x} \} \big)
} 
\;\; \nrule{[Join]}
\qquad
\irule{P \;\equiv\; P' \rightarrow Q' \;\equiv\; Q}{
P \rightarrow Q
}
\;\; \nrule{[Struct]}
\end{array}
\]
\caption{The reduction relation} \label{fig:reduction}
\end{figure}

We now comment the rules for reduction. 
Rule \nrule{Tau} simply fires the $\tau$ prefix. 
Rule \nrule{Tell} augments the context ($R$) with a constraint $c$. 
Similarly to~\cite{Saraswat91cc}, we do not check for the consistency 
of~$c$ with the other constraints in~$R$. 
If desired, a side condition similar to that of rule \nrule{Check} 
(discussed below) can be added, at the cost reduced compositionality. 
As another option, one might restrict the constraint $c$ in $\tell c.P$ to a
class of coherent constraints, as done e.g.~in~\cite{BZ10lics}.  
% Back to our rules. 
Rule \nrule{Ask} checks whether the context has enough
active constraints $C$ so to entail $c$. 
Rule \nrule{Check} checks the context for consistency with $c$.
Since this requires inspecting every active constraint in the context, 
a side condition precisely separates the context between $C$ and $R$, 
so that all the active constraints are in $C$,
which in this case acts as a global {\em constraint store}.

Rule \nrule{Fuse} replaces a set of variables $x\vec{y}$ with a
bound name $n$, hence fusing all the variables together. 
One variable in the set, $x$, is the one mentioned in the $\fuse{x}{c}$ prefix,
while the others, $\vec{y}$, are taken from the context. 
The replacement of variables is done by the substitution $\sigma$
in the rule premises.
The actual set of variables $\vec{y}$ to fuse is chosen according to
the {\em minimal fusion} policy, formally defined below.
% We require that $\sigma$ is a {\em minimal fusion} for $C \vdash c$, as 

\begin{definition}[Minimal Fusion] \label{def:minimal-fusion}
A fusion $\sigma = \setenum{\bind{\vec{z}}{n}}$ is 
\emph{minimal} for $C, c$,
written \mbox{$C \vdash^{min}_{\sigma} c$}, iff:
\[
  C \sigma \vdash c \sigma
  \;\;\land\;\;
  \nexists \, \vec{w}\subsetneq \vec{z} \;\, : \;
  C \{\nicefrac{n}{\vec{w}}\} \vdash c \{\nicefrac{n}{\vec{w}}\}
\]
\end{definition}

A minimal fusion $\sigma$ must cause the entailment of $c$ by the context $C$. 
Furthermore, fusing a proper subset of variables must not cause the entailment.
The rationale for minimality is that we want to fuse
those variables only, which are actually involved in the entailment
of~$c$ -- not any arbitrary superset.  
Pragmatically, we will often use $\fuse xc$ as a construct to establish 
{\em sessions}: the participants are then chosen among those actually involved 
in the satisfaction of the constraint $c$, and each participant ``receives''
the fresh name $n$ through the application of $\sigma$. 
In this case, $n$ would act as a sort of {\em session identifier}.

Note that the context $R$ in rule \nrule{Fuse} may contain active constraints. 
So, the fusion $\sigma$ is actually required to be minimal with respect to 
a {\em subset} $C$ of the active constraints of the whole system. 
Technically, this will allow us to provide a compositional semantics 
for $\fuse{x}{c}$. 
Also, this models the fact that processes have a ``local'' 
view of the context, as we will discuss later in this section.

Rule \nrule{Join} replaces a variable $x$ 
with a name $n$ taken from the context. 
Note that, unlike \nrule{Fuse}, $n$ is \emph{not} fresh here.  
To enable a $\join{x}{c}$ prefix, the substitution must cause~$c$
to be entailed by the context $C$. 
Intuitively, this prefix allows to ``search'' in the context 
for some $x$ satisfying a constraint $c$.
This can also be used to join a session which was previously
established by a \nrule{Fuse}.

Note that rules \nrule{Fuse} and \nrule{Join} provide a
non-deterministic semantics for prefixes $\fuse{}{}$ and $\join{}{}$
since several distinct fusions $\sigma$ could be used to derive a
transition.  Each $\sigma$ involves only names and variables occurring
in the process at hand, plus a fresh name $n$ in the case of
$\fuse{}{}$. If we consider $n$ up-to renaming, we have a finite
number of choices for $\sigma$.  Together with guarded recursion, this
makes the transition system to be finitely-branching.

Rule \nrule{Struct} simply allows us to consider processes up-to
structural equivalence.

\vspace{-2pt}
\paragraph{Transition semantics.}
We now present an alternative semantics, specified
through a labelled transition relation. 
Unlike the reduction semantics, the labelled relation
$\xrightarrow{\alpha}$ is {\em compositional}: all the prefixes can be
fired by considering the relevant portion of the system at hand.  The
only exception is the $\checkp c$ prefix, which is inherently
non-compositional. We deal with $\checkp c$ by layering the reduction
relation $\rightarrowtail$ over the relation $\xrightarrow{\alpha}$.
While defining the transition semantics, we borrow most of the intuitions
from the semantics in~\cite{BZ10lics}.
The crucial difference between the two is how they
finalize the actions generated by a $\fuse{}{}$ (rule \nrule{CloseFuse}).
Roughly, in~\cite{BZ10lics} we need a quite complex treatment, since there 
we have to accommodate with ``principals'' mentioned in the constraints.
Since here we do not consider principals, we can give a smoother treatment.
We discuss in detail such issues in Sect.~\ref{sec:related-work}.

We start by introducing in Def.~\ref{def:actions} the \emph{actions} of
our semantics, that is the set of admissible labels of the LTS.
The transition relation is then presented in Def.~\ref{def:lts}.

\begin{definition}[Actions]\label{def:actions}
Actions $\alpha$ are as follows, where $C$ denotes a set of constraints.
\begin{align*}
\alpha & \; \bnfdef \; \tau \;\bnfmid\;
            C \;\bnfmid\;
            C\vdash c \;\bnfmid\;
            C\vdash_{x}^F c \;\bnfmid\;
            C\vdash_{x}^J c \;\bnfmid\;
            C\not\vdash\bot \;\bnfmid\;
            (a) \, \alpha
\end{align*}
\end{definition}

The action $\tau$ represents an internal move. 
The action $C$ advertises a set of active constraints.
% The action $C$ is generated when a process advertises the
% set of active constraints it carries.
% (without actually moving).
The action $C\vdash c$ is a {\em tentative} action, 
generated by a process attempting to fire an $\ask c$ prefix. 
This action carries the collection $C$
of the active constraints discovered so far. 
Similarly for $C\vdash_{x}^F c$ and $\fuse{x}{c}$, 
for $C\vdash_{x}^J c$ and $\join{x}{c}$,
as well as for $C\not\vdash\bot$ and $\checkp c$. 
In the last case, $C$ includes $c$.
The delimitation in
$(a)\alpha$ is for scope extrusion,
as in the labelled semantics of the $\pi$-calculus~\cite{Sangiorgi01pi}. 
We write $(\vec{a})\alpha$ to denote a set of distinct delimitations, 
neglecting their order, e.g.\ $(ab)=(ba)$. 
We simply write $(\vec{a}\vec{b})$ for $(\vec{a}\cup\vec{b})$.

\begin{definition}[Transition relation] \label{def:lts}
The transition relations $\xrightarrow{\alpha}$ 
are the smallest relations between processes 
satisfying the rules in Fig.~\ref{fig:lts}.
The last two rules in Fig.~\ref{fig:lts} define 
the reduction relation~$\rightarrowtail$.
% layered over the $\xrightarrow{\alpha}$ relations.
% through a two-layered transition system
%The bottom layer is an LTS $\xrightarrow{\alpha}$ between processes,
%which provides a compositional semantics.
\end{definition}

\begin{figure*}[t!]
\small
\[
\begin{array}{c}
  \tau.P \xrightarrow{\tau} P 
  \;\;\; 
  \nrule{[Tau]}
\qquad
  \ask c.P\xrightarrow{\emptyset\vdash c}P 
  \;\;\; 
  \nrule{[Ask]}
\qquad
  \tell c.P\xrightarrow{\tau} c \, | \, P 
  \;\;\; 
  \nrule{[Tell]}
\\[8pt]
  \checkp c.P\xrightarrow{\setenum{c}\not\vdash\bot}P 
  \;\;\; 
  \nrule{[Check]} 
\qquad
   \fuse{x}{c}.P\xrightarrow{\emptyset\vdash_{x}^F c}P
   \;\;
   \nrule{[Fuse]}
\qquad
   \join{x}{c}.P\xrightarrow{\emptyset\vdash_{x}^J c}P
   \;\;
   \nrule{[Join]}
\\[8pt]
   u\xrightarrow{\{u\}}u
\;\;\;
   \nrule{[Constr]}
\qquad
   \sum_i \pi_i.P_i \xrightarrow{\emptyset} \sum_i \pi_i.P_i
   \;\;
   \nrule{[IdleSum]}
\\[8pt]
   \irule{P\xrightarrow{(\vec{a})C}P'\,\,\,\, Q\xrightarrow{(\vec{b})C'}Q'}{P|Q\xrightarrow{(\vec{a}\vec{b})(C\cup C')}P'|Q'} \;
\nrule{[ParConstr]}
\quad
   \irule{P\xrightarrow{(\vec{a})C}P'\,\,\,\, Q\xrightarrow{(\vec{b})(C'\vdash c)}Q'}{P|Q\xrightarrow{(\vec{a}\vec{b})(C\cup C'\vdash c)}P'|Q'}\;
\nrule{[ParAsk]}
\\[20pt]
   \irule{P\xrightarrow{(\vec{a})C}P'\,\,\,\, Q\xrightarrow{(\vec{b})(C'\vdash_{x}^F c)}Q'}{P|Q\xrightarrow{(\vec{a}\vec{b})(C\cup C'\vdash_{x}^F c)}P'|Q'}\;
\nrule{[ParFuse]} 
\quad
  \irule{P\xrightarrow{(\vec{a})C}P'\,\,\,\, Q\xrightarrow{(\vec{b})(C'\vdash_{x}^{J}c)}Q'}{P|Q\xrightarrow{(\vec{a}\vec{b})(C\cup C'\vdash_{x}^{J}c)}P'|Q'}\;\;
\nrule{[ParJoin]}
\\[20pt]
  \irule{P\xrightarrow{(\vec{a})C}P'\,\,\,\, Q\xrightarrow{(\vec{b})(C'\not\vdash\bot)}Q'}{P|Q\xrightarrow{(\vec{a}\vec{b})(C\cup C'\not\vdash\bot)}P'|Q'}\;\;
\nrule{[ParCheck]}
\qquad
  \irule{P\xrightarrow{\tau} P'}{P|Q'\xrightarrow{\tau} P'|Q'} \;\;
  \nrule{[ParTau]}
\\[20pt]
   \irule{\pi_j.P_j\xrightarrow{\alpha}P'}{\sum_i \pi_i.P_i \xrightarrow{\alpha}P'}
   \;
   \nrule{[Sum]}
\quad\;\;
   \irule{P\{\nicefrac{\vec{a}}{\vec{x}}\}\xrightarrow{\alpha}P'}{X(\vec{a)}\xrightarrow{\alpha}P'}\mbox{ if }X(\vec{x})\doteq P
   \;
   \nrule{[Def]}
\quad\;\;
   \irule{P\xrightarrow{\alpha}P'}{(a)P\xrightarrow{\alpha}(a)P'}\mbox{ if }a\not\in\alpha
   \; 
   \nrule{[Del]}
\\[15pt]
   \irule{P\xrightarrow{\alpha}P'}{(a)P\xrightarrow{(a)\alpha}P'}
   \;
   \nrule{[Open]}
\quad\;\;
   \irule{P\xrightarrow{(\vec{a})(C\vdash c)}P'}{P\xrightarrow{\tau}(\vec{a})P'}\mbox{ if }C\vdash c
   \;
   \nrule{[CloseAsk]}
\\[20pt]
   \irule{
     P\xrightarrow{(x\vec{y}\vec{a})(C \,\vdash^F_{x}\, c)}P' \quad
     \sigma = \setenum{\bind{x\vec{y}}{n}} \quad
     \mbox{$n$ fresh} \quad
     C \vdash_{\sigma}^{loc} c
   }
   {P\xrightarrow{\tau}(n\vec{a})(P'\sigma)}
\;\; \nrule{[CloseFuse]}
\\[15pt]
   \irule{P\xrightarrow{(xn\vec{a})(C\vdash_{x}^{J}c)}P'}{P\xrightarrow{\tau}(n\vec{a})P'\sigma}\mbox{ if } \begin{array}{l} C\sigma\vdash c\sigma, \\ \sigma=\{\nicefrac{n}{x} \} \end{array}
   \;
   \nrule{[CloseJoin]}
\qquad
   \irule{P\xrightarrow{\tau} P'}{P\rightarrowtail P'}
   \; \nrule{[TopTau]}
\\[20pt]
   \irule{P\xrightarrow{(\vec{a})(C\not\vdash\bot)}P'}{P\rightarrowtail(\vec{a})P'}\mbox{ if } C\not\vdash\bot
   \;
   \nrule{[TopCheck]}
\end{array}
\]
\caption{The labelled transition relation.
Symmetric rules for $+,|$ are omitted. 
The rules \nrule{Par*} have the following no-capture side condition:
$\vec{a}$ is fresh in $\vec{b},C',c,x,Q'$, 
while $\vec{b}$ is fresh in $C,P'$.}
\label{fig:lts}
\end{figure*}

Many rules in Fig.~\ref{fig:lts} are rather standard, so we comment on the
most peculiar ones, only. 
Note in passing that $\equiv$ is not used in this semantics.
The rules for prefixes simply generate the
corresponding tentative actions. Rule \nrule{Constr} advertises an
active constraint, which is then used to augment the tentative actions
through the \nrule{Par}* rules. Rule \nrule{Open} lifts a restriction
to the label, allowing for scope extrusion. The \nrule{Close}* rules
put the restriction back at the process level, and also convert
tentative actions into $\tau$. 

The overall idea is the following: a tentative action label carries
all the proof obligations needed to fire the corresponding prefix.
The \nrule{Par}*
rules allow for exploring the context, and augment the label with the
observed constraints. The \nrule{Close}* rules check that enough
constraints have been collected so that the proof obligations can
be discharged, and transform the label into a $\tau$.

The \nrule{Top}* rules act on the top-level, only, and define the
semantics of $\checkp{c}$.

The side condition of rule \nrule{CloseFuse} involves a variant of the
minimal fusion relation we used previously. As for the reduction
semantics, we require $\sigma$ to be minimal, so not to fuse more
variables than necessary. Recall however that in the reduction
semantics minimality was required with respect to a {\em part} of the
active constraints at hand. In our labelled semantics, rules
\nrule{Par}* collect each active constraint found in the syntax tree
of the process. If we simply used $C \vdash^{min}_\sigma c$ in
\nrule{CloseFuse}, we would handle the following example differently.
Let
$C = \mathsf{q}(y) \, | \, \mathsf{q}(z) \lor \mathsf{s} \imp \mathsf{p}(y)$,
let $P = (x)(y)(z) ( \fuse{x}{\mathsf{p}(x)}.P \mid C \mid \mathsf{s} )$, and
let $Q = (x)(y)(z)(\fuse{x}{\mathsf{p}(x)}.P \mid C) \mid \mathsf{s}$.
In $P$ we must collect $s$ before applying \nrule{CloseFuse}, and so
$\sigma_1=\{\bind{xy}{n}\}$ would be the only minimal fusion.
Instead in $Q$ we can also apply \nrule{CloseFuse} before discovering $s$,
yielding the minimal fusion $\sigma_2=\{\bind{xyz}{n}\}$. This would
be inconsistent with $\equiv$ (and our reduction semantics as well).
To recover this, we instead require in \nrule{CloseFuse} the following
relation, stating that $\sigma$ must be minimal with respect to a part
of the observed constraints, only.

\begin{definition}[Local Minimal Fusion] \label{def:local-minimal-fusion}
A fusion $\sigma = \setenum{\bind{\vec{z}}{n}}$ is 
\emph{local minimal} for $C, c$,
written \mbox{$C \vdash^{loc}_{\sigma} c$}, \nolinebreak iff:
\[
  \exists C' \subseteq C \;\, : \; C' \vdash^{min}_{\sigma} c
\]
\end{definition}

While we did not use structural equivalence in the definition of the
labelled transition semantics, it turns out to be a bisimulation.

\begin{theorem} The relation $\equiv$ is a bisimulation, \text{i.e.}:
\(
  P \equiv Q \xrightarrow{\alpha} Q'  \implies 
  \exists P'.\; P \xrightarrow{\alpha} P' \equiv Q'
\).
\end{theorem}

We also have the expected correspondence between the reduction and
labelled semantics.

\begin{theorem}
\(
  P \rightarrow P' \iff \exists Q,Q'.\, P \equiv Q \rightarrowtail Q' \equiv P'
\)
\end{theorem}

The right implication is by rule induction. 
To prove the left implication, an induction argument on 
$Q \xrightarrow{\alpha} Q'$ suffices,
exploiting the fact that all the constraints of~$Q$ 
are accumulated in the label.

\section{Examples} \label{sec:examples}

We illustrate our calculus by modelling scenarios where the interaction
among parties is driven by contracts.
In all the examples below, we use as constraints a smooth extension
of the propositional contract logic \pcl\!~\cite{BZ10lics}.
A comprehensive presentation of \pcl is beyond the scope of this paper,
so we give here just a broad overview, and we refer the reader 
to~\cite{BZ10lics,PCLtr} for all the technical details and further examples.

\pcl is an extension of intuitionistic propositional logic IPC~\cite{Troelstra},
featuring a \emph{contractual implication} connective~$\coimp$.
Differently from IPC,
a ``contract'' $\sf b \coimp a$ implies $\sf a$ not only when $\sf b$ is true, 
like IPC implication, but also in the case that a ``compatible'' 
contract, e.g.\ $\sf a \coimp b$, holds.
%That is,
%$\vdash \sf (b \coimp a) \;\land\; (a \coimp b) \;\imp\; a \land b$
%is a theorem in \pcl\!.
So, \pcl allows for sort of ``circular'' assume-guarantee reasoning,
summarized by the theorem 
$\vdash \sf (b \coimp a) \;\land\; (a \coimp b) \;\imp\; a \land b$.

The proof system of \pcl extends that of IPC
with the following axioms:
\begin{align*}
& \top \coimp \top 
& (p \coimp p) \imp p
&& (p' \imp p) \imp (p \coimp q) \imp (q \imp q') \imp (p' \coimp q')
\end{align*}

A main result about \pcl is its decidability, proved via cut elimination.
%Below, we allow prime formulae of the form $\mathsf{p}(\vec{a})$.
%Note that parameters $\vec{a}$ still keep \pcl a propositional logic. 
Therefore, we can use the (decidable) provability relation of \pcl
as the entailment relation $\vdash$ of the constraint structure.

\begin{example}[Greedy handshaking]\label{ex:proc:handshaking}

Suppose there are three kids who want to play together.
Alice has a toy airplane, Bob has a bike, while Carl has a toy car.
Each of the kids is willing to share his toy, but only provided that
the other two kids promise they will lend their toys to him.
So, before sharing their toys, 
the three kids stipulate a ``gentlemen's agreement'',
modelled by the following \pcl contracts:
\begin{small}
\[
  c_{\it Alice}(x) \,=\, ({\sf b}(x) \land {\sf c}(x)) \coimp {\sf a}(x) 
  \qquad 
  c_{\it Bob}(y) \,=\, ({\sf a}(y) \land {\sf c}(y)) \coimp {\sf b}(y) 
  \qquad
  c_{\it Carl}(z) \,=\, ({\sf a}(z) \land {\sf b}(z)) \coimp {\sf c}(z)
\]
\end{small}
Alice's contract $c_{\it Alice}(x)$ says that
Alice promises to share her airplane in a session $x$, written ${\sf a}(x)$, 
provided that both Bob and Carl will share their toys in the same session. 
Bob's and Carl's contracts are dual.
The proof system of \pcl allows to deduce that the three kids will
indeed share their toys in any session $n$, i.e.\ 
$c_{\it Alice}(n) \,\land\, c_{\it Bob}(n) \,\land\, c_{\it Carl}(n)
\;\imp\; {\sf a}(n) \,\land\, {\sf b}(n) \,\land\, {\sf c}(n)$
is a theorem of \pcl\!.
% \begin{small}
% \begin{equation*}
%   c_{\it Alice}(n) \,\land\, c_{\it Bob}(n) \,\land\, c_{\it Carl}(n)
%   \;\;\imp\;\; 
%   {\sf a}(n) \,\land\, {\sf b}(n) \,\land\, {\sf c}(n)
% \end{equation*}
% \end{small}
%All the above shows how to deduce, from a set of contracts, 
%the rights and the duties of the involved parties.
We model the actual behaviour of the three kids through the following processes:
% \begin{align*}
%   \mathit{Alice} & = 
%   (x) \; \big( \tell{c_{\it Alice}(x)}.\; \fuse{x}{\mathsf{a}(x)}.\, \mathit{lendA} \big)
% \\
%   \mathit{Bob} & = 
%   (y) \; \big( \tell{c_{\it Bob}(y)}.\; \fuse{y}{\mathsf{b}(y)}.\, \mathit{lendB} \big) \\
%   \mathit{Carl} & = 
%   (z) \; \big( \tell{c_{\it Carl}(z)}.\; \fuse{z}{\mathsf{c}(z)}.\, \mathit{lendC} \big)
% \end{align*}
\[
\small
\begin{array}{c}
  \mathit{Alice} = 
  (x) \; \big( \tell{c_{\it Alice}(x)}.\; \fuse{x}{\mathsf{a}(x)}.\, \mathit{lendA} \big)
\qquad
  \mathit{Bob} = 
  (y) \; \big( \tell{c_{\it Bob}(y)}.\; \fuse{y}{\mathsf{b}(y)}.\, \mathit{lendB} \big)
\\[5pt]
  \mathit{Carl} = 
  (z) \; \big( \tell{c_{\it Carl}(z)}.\; \fuse{z}{\mathsf{c}(z)}.\, \mathit{lendC} \big)
\end{array}
\]
A possible trace of the LTS semantics is the following:
\begin{small}
\begin{align*}
 & \mathit{Alice} \mid \mathit{Bob} \mid \mathit{Carl} 
 \xrightarrow{\tau} \  (x) \; \big( c_{\it Alice}(x) \;|\; \fuse{x}{\mathsf{a}(x)}.\, \mathit{lendA} \big) \mid {\it Bob} \mid {\it Carl} \\
 \xrightarrow{\tau}\ & (x) \; \big( c_{\it Alice}(x) \;|\; \fuse{x}{\mathsf{a}(x)}.\, \mathit{lendA} \big) \; | \;
 (y) \; \big( c_{\it Bob}(y) \;|\; \fuse{y}{\mathsf{b}(y)}.\, \mathit{lendB} \big) \mid {\it Carl} \\
 \xrightarrow{\tau}\ & (x) \; \big( c_{\it Alice}(x) \;|\; \fuse{x}{\mathsf{a}(x)}.\, \mathit{lendA} \big) \; | \;
 (y) \; \big( c_{\it Bob}(y) \;|\; \fuse{y}{\mathsf{b}(y)}.\, \mathit{lendB} \big) \mid  
 (z) \; \big( c_{\it Carl}(z) \;|\; \fuse{z}{\mathsf{c}(z)}.\, \mathit{lendC} \big) \\
 \xrightarrow{\tau}\ & (n) \; \big( c_{\it Alice}(n) \;|\; \mathit{lendA}\{\nicefrac{n}{x}\} \; | \;
 c_{\it Bob}(n) \;|\; \ask{\mathsf{b}(n)}.\, \mathit{lendB}\{\nicefrac{n}{y}\}  \; | \;
 c_{\it Carl}(n) \;|\; \ask{\mathsf{c}(n)}.\, \mathit{lendC}\{\nicefrac{n}{y}\} \big) \\
 \xrightarrow{\tau}\ & (n) \; \big( c_{\it Alice}(n) \;|\; \mathit{lendA}\{\nicefrac{n}{x}\} \; | \;
 c_{\it Bob}(n) \;|\; \mathit{lendB}\{\nicefrac{n}{y}\} \; | \;
 c_{\it Carl}(n) \;|\; \ask{\mathsf{c}(n)}.\, \mathit{lendC}\{\nicefrac{n}{y}\} \big)
 \\
 \xrightarrow{\tau}\ & (n) \; \big( c_{\it Alice}(n) \;|\; \mathit{lendA}\{\nicefrac{n}{x}\} \; | \;
 c_{\it Bob}(n) \;|\; \mathit{lendB}\{\nicefrac{n}{y}\} \; | \;
 c_{\it Carl}(n) \;|\; \mathit{lendC}\{\nicefrac{n}{y}\} \big)
\end{align*}
\end{small}
In step one, we use \nrule{Tell,ParTau,Del} to fire the prefix
$\tell{c_{\it Alice}(x)}$. 
Similarly, in steps two and three, 
we fire the prefixes $\tell{c_{\it Bob}(y)}$ and $\tell{c_{\it Carl}(z)}$.
Step four is the crucial one. 
There, the prefix $\fuse{x}{\mathsf{a}(x)}$ is fired through rule \nrule{Fuse}.
Through rules \nrule{Constr,ParFuse}, we discover the active constraint 
$c_{\it Alice}(x)$. 
We use rule \nrule{Open} to obtain the action 
$(x) \{ c_{\it Alice}(x) \}\vdash^F_x \mathsf{a}(x)$ for the Alice part. 
For the Bob part, we use rule \nrule{Constr} to discover 
$c_{\it Bob}(y)$, which we then merge with the
empty set of constraints obtained through rule \nrule{IdleSum}; 
similarly for Carl.
We then apply \nrule{Open} twice and obtain $(y) \{ c_{\it Bob}(y) \}$ and
$(z) \{ c_{\it Carl}(z) \}$. 
At the top level, we apply \nrule{ParFuse} to deduce 
$(x,y,z) \{ c_{\it Alice}(x),c_{\it Bob}(y),c_{\it Carl}(z)\} \vdash^F_x \mathsf{a}(x)$. 
Finally, we apply \nrule{CloseFuse}, which fuses $x$, $y$ and $z$ 
by instantiating them to the fresh name $n$. 
It is easy to check that 
$\{ c_{\it Alice}(x), c_{\it Bob}(y), c_{\it Carl}(z)\}\vdash^{loc}_{\sigma} \mathsf{a}(x)$ where $\sigma = \{\bind{xyz}{n} \}$. 
Note that all the three kids have to cooperate, in order to proceed.
Indeed, fusing e.g.~only $x$ and $y$ would not allow
to discharge the premise $c(z)$ from the contracts $c_{\it Alice}$ and
$c_{\it Bob}$, hence preventing any $\fuse{}{}$ prefix from being fired.

%Note that the instantiation causes $\fuse{y}{\mathsf{b}(y)}$ 
%to transform into $\ask{\mathsf{b}(n)}$,
%so that it can be fired in the last step.
\end{example}

\begin{example}[Insured Sale]
A seller $S$ will ship an order as long as she is either paid upfront, 
or she receives an insurance from the insurance company $I$, which she trusts.
We model the seller contract as the \pcl formula 
$s(x) =  
  \mathsf{order}(x) \land (\mathsf{pay}(x) \lor \mathsf{insurance}(x))
  \, \coimp \,
  \mathsf{ship}(x)$
where $x$ represents the session where the order is placed. 
The seller $S$ is a recursive process, allowing multiple
orders to be shipped.
\begin{align*}
  S & \; \doteq \; (x) \, \tell {s(x)}.\, \fuse{x}{\mathsf{ship}(x)}.\, (S \mid \mathit{doShip}(x))
\end{align*}
The insurer contract 
$i(x) = \mathsf{premium}(x) \coimp \mathsf{insurance}(x)$
plainly states that a premium must be paid upfront.
The associated insurer process $I$ is modelled as follows:
\[
  I \doteq (x) \tell i(x). \fuse{x}{\mathsf{insurance}(x)}.\;
   \big( I \, |\, 
   \tau . \checkp {\lnot \mathsf{pay}(x)} . (\mathit{refundS}(x) \mid \mathit{debtCollect}(x)) \big)
\]
When the insurance is paid for, the insurer will wait for some time,
modelled by the $\tau$ prefix. 
After that, he will check whether the buyer has not paid the shipped goods. 
In that case, 
the insurer will immediately indemnify the seller, and contact a debt
collector to recover the money from the buyer.
Note that~$S$ and~$I$ do not explicitly mention any specific buyer. 
As the interaction among the parties is loosely specified, 
many scenarios are possible. 
For instance, consider the following buyers $B_0,B_1,B_2,B_3$:
\begin{small}
\begin{align*}
b_0(x) & \;=\; \mathsf{ship}(x) \coimp \mathsf{order}(x) \land 
        \mathsf{pay}(x) 
&
B_0 & \;=\; (x) \, \tell b_0(x). \mathit{receive}(x) \\
b_1(x) & \;=\; \mathsf{ship}(x) \coimp \mathsf{order}(x) \land 
        \mathsf{premium}(x)
&
B_1 & \;=\; (x) \, \tell b_1(x). (\mathit{receive}(x) \mid \tau. \tell \mathsf{pay}(x) ) \\
b_2(x) & \;=\; \mathsf{order}(x) \land \mathsf{pay}(x) 
& 
B_2 & \;=\; B_0\{\nicefrac{b_2}{b_0}\} \\
b_3(x) & \;=\; \mathsf{order}(x) \land \mathsf{premium}(x)
&
B_3 & \;=\; B_0\{\nicefrac{b_3}{b_0}\}
\end{align*}
\end{small}
The buyer $B_0$ pays upfront.
The buyer $B_1$ will pay later, by providing the needed insurance.
The ``incautious'' buyer $B_2$ will pay upfront, 
without asking any shipping guarantees.
The buyer $B_3$ is insured, and will not pay.
The insurer will then refund the seller, and start a debt collecting
procedure. This is an example where a violated promise can be detected
so to trigger a suitable recovery action.
%
% Summing up, note the role of the \nrule{CloseFuse} rule in
% these scenarios: the minimality requirement makes sure the insurer is
% involved only when actually needed. 
The minimality requirement guarantees that the insurer will be
involved only when actually needed. 
\end{example}

\begin{example}[Automated Judge]\label{ex:proc:judge}
Consider an online market, where buyers and sellers trade items. 
The contract of a buyer is to pay for an item, 
provided that the seller promises to send it; dually, 
the contract of a seller is to send an item, provided that the buyer pays.
A buyer first issues her contract, then waits until discovering she
has to pay, and eventually proceeds with the process $\mathit{CheckOut}$.
At this point, 
the buyer may either abort the transaction (process $\mathit{NoPay}$),
or actually pay the item, by issuing the constraint $\mathsf{paid}(x)$.
After the item has been paid, the buyer may wait for the item to be sent
($\ask{\mathsf{sent}(x)}$),
or possibly open a dispute with the seller ($\tell{\mathsf{dispute}(x)}$).
Note that, as in the real world, one can always open a dispute, 
even when the other party is perfectly right.
\begin{align*}
  \mathit{Buyer} & = 
  (x) \; \big( \tell{\,\mathsf{send}(x) \coimp \mathsf{pay}(x)}.\; \fuse{x}{\mathsf{pay}(x)}.\, \mathit{CheckOut} \big) \\
  \mathit{CheckOut} & = \tau.\mathit{NoPay} + \tau.\,\tell{\mathsf{paid}(x)}. (\tau.\,\tell{\mathsf{dispute}(x)} + \ask{\mathsf{sent}(x)})
\end{align*}
The behaviour of the seller is dual:
issue the contract, wait until she has to send, 
and then proceed with $\mathit{Ship}$.
There, either choose not to send, or send the item and then 
wait for the payment or open a dispute.
\begin{align*}
\mathit{Seller} & = 
  (y) \; \big( \tell{\mathsf{pay}(y) \coimp \mathsf{send}(y)}.\; \fuse{y}{\mathsf{send}(y)}.\, \mathit{Ship} \big) \\
  \mathit{Ship} & = \tau.\mathit{NoSend} + \tau.\,\tell{\mathsf{sent}(y)}. (\tau.\, \tell{\mathsf{dispute}(y)} + \ask{\mathsf{paid}(y)})
\end{align*}
To automatically resolve disputes, the process $\mathit{Judge}$ may enter
a session initiated between a buyer and a seller, provided that a dispute
has been opened, and either the obligations $\mathsf{pay}$ or $\mathsf{send}$
have been inferred.
This is done through the $\join{z}{}$ primitive, which binds the variable~$z$
to the name of the session established between buyer and seller.
If the obligation $\mathsf{pay}(z)$ is found but the item has not been
paid (i.e.~$\checkp{\neg\mathsf{paid}(z)}$ passes),
then the buyer is convicted 
(by~$\mathit{jailBuyer(z)}$, not further detailed).
Similarly, if the obligation $\mathsf{send}(z)$ has not been supported by 
a corresponding $\mathsf{sent}(z)$, the seller is convicted.
\begin{align*}
\mathit{Judge} & = 
  (z) \; \big( 
  \join{z}{(\mathsf{pay}(z) \land \mathsf{dispute}(z))}. \checkp{\neg\mathsf{paid}(z).\, \mathit{jailBuyer(z)}} \mid \\
  & \phantom{= (z) \; \big( \;} 
  \join{z}{(\mathsf{send}(z) \land \mathsf{dispute}(z))}. \checkp{\neg\mathsf{sent}(z).\, \mathit{jailSeller(z)}} 
  \big)
\end{align*}
A possible trace of the LTS semantics is the following:
\begin{small}
\begin{align*}
 & \mathit{Buyer} \!\mid\! \mathit{Seller} \!\mid\! \mathit{Judge} 
 \xrightarrow{\tau}^*  (n) \; \big( \mathsf{send}(n) \coimp \mathsf{pay}(n) 
 \!\mid\! \mathsf{paid}(n) \!\mid\! \tell{\mathsf{dispute}(n)} \!\mid\! 
 \mathsf{pay}(n) \coimp \mathsf{send}(n) \!\mid\! \mathit{NoSend} \big) \!\mid\! \mathit{Judge} \\
&\xrightarrow{\tau}^* (n) \; \big( \mathsf{send}(n) \coimp \mathsf{pay}(n) 
 \!\mid\! \mathsf{paid}(n) \!\mid\! \mathsf{dispute}(n) \!\mid\! 
 \mathsf{pay}(n) \coimp \mathsf{send}(n) \!\mid\! \mathit{NoSend} \!\mid\!
 \checkp{\neg\mathsf{sent}(n)}.\, \mathit{jailSeller(n)} \!\mid\! \cdots \big) \\
& \xrightarrow{\tau}^* (n) \; \big( \mathit{jailSeller(n)} \!\mid\! \cdots \big)
\end{align*}
\end{small}
A more complex version of this example is in~\cite{BZ10lics}, also dealing
with the identities of the principals performing the relative promises.
The simplified variant presented here does not require the 
more general rule for $\fuse{}{}$ found in~\cite{BZ10lics}.
\end{example}

\begin{example}[All-you-can-eat]\label{ex:oplus}
Consider a restaurant offering an all-you-can-eat buffet.
Customers are allowed to have a single trip down the buffet line, 
where they can pick anything they want. 
After the meal is over, they are no longer allowed to return to the buffet.
% and eat the dishes they did not eat before.
In other words, multiple dishes can be consumed, but only in a single step.
We model this scenario as follows:
% \begin{align*}
% {\it Buffet} & = (x) \; ( \mathsf{pasta}(x) \mid \mathsf{chicken}(x) \mid
% \mathsf{cheese}(x) \mid \mathsf{fruit}(x) \mid \mathsf{cake}(x) ) \\
% {\it Bob}  & = (x) \; \fuse{x}{\mathsf{pasta}(x) \land \mathsf{chicken}(x)}.\, {\it SatiatedB} \\
% {\it Carl} & = (x) \; \fuse{x}{\mathsf{pasta}(x)} . \fuse{x}{\mathsf{chicken}(x)}.\, {\it SatiatedC}
% \end{align*}
\vspace{-2pt}
\[
\small
\begin{array}{c}
 {\it Buffet} = (x) \; ( \mathsf{pasta}(x) \mid \mathsf{chicken}(x) \mid
 \mathsf{cheese}(x) \mid \mathsf{fruit}(x) \mid \mathsf{cake}(x) ) \\[8pt]
 {\it Bob}  = (x) \; \fuse{x}{\mathsf{pasta}(x) \land \mathsf{chicken}(x)}.\, {\it SatiatedB} 
 \quad
 {\it Carl} = (x) \; \fuse{x}{\mathsf{pasta}(x)} . \fuse{x}{\mathsf{chicken}(x)}.\, {\it SatiatedC}
\end{array}
\]
The Buffet can interact with either Bob or Carl, and make them satiated.
Bob eats both pasta and chicken in a single meal,
while Carl eats the same dishes but in two different meals, 
thus violating the Buffet policy, \text{i.e.}:
\(
  {\it Buffet} \mid {\it Carl} 
  \; \rightarrow^* \;
  {\it SatiatedC} \mid P
\).
Indeed, the Buffet should forbid Carl to
eat the chicken, i.e.~to fire the second $\fuse{x}{}$.  
To enforce the Buffet policy, we first define the auxiliary operator~$\oplus$. 
Let $(p_i)_{i \in I}$ be \pcl formulae,
let $\mathsf{r}$ be a fresh prime,
$o$ a fresh name, and $z,(z_i)_{i\in I}$ fresh variables.
Then:
\vspace{-4pt}
\[
\small
\textstyle \bigoplus_{i\in I} p_i
= (o)(z)(z_i)_{i\in I}(\mathsf{r}(o,z) \mid
               |\! |_{i\in I} \mathsf{r}(o,z_i)\imp p_i)
\]
To see how this works, consider the process $\oplus_{i\in I}\, p_i | Q$
where $Q$ fires a $\fuse{x}{}$ which demands a subset of the 
constraints $(p_i)_{i\in J}$ with $J \subseteq I$.
To deduce $p_i$ we are forced to fuse $z_i$ with $z$ (and $x$); otherwise
we can not satisfy the premise $\mathsf{r}(o,z_i)$. Therefore all the
$(z_i)_{i\in J}$ are fused, while minimality of fusion ensures that the
$(z_i)_{i\in I\setminus J}$ are not.
After fusion we then reach:
\vspace{-4pt}
\[
\small
(o)(m) \big( (z_i)_{i\in I\setminus J}
    (\, |\! |_{i\in I\setminus J} \, \mathsf{r}(o,z_i)\imp p_i )
|\, |\! |_{i\in J} ( \mathsf{r}(o,m) | \mathsf{r}(o,m)\imp p_i ) 
   \big)
\,\mid\, Q'
\]
where $m$ is a fresh name resulting from the fusion.
Note that the $(p_i)_{i\in I\setminus J}$ can no longer be deduced 
through fusion, since the variable $z$ was ``consumed'' by the first 
fusion. 
The rough result is that
$\oplus_i \, p_i$ allows a subset of the  $(p_i)_{i\in I}$ to be demanded
through fusion, after which the rest is no longer available.

We can now exploit the $\oplus$ operator to redefine the Buffet as follows:
\[
\small
  {\it Buffet'} = (x) (\mathsf{pasta}(x) \oplus \mathsf{chicken}(x) \oplus
  \mathsf{cheese}(x) \oplus \mathsf{fruit}(x) \oplus \mathsf{cake}(x) ) \\
\]
The new specification actually enforces the Buffet policy, \text{i.e.}:
\(
  {\it Buffet'} \mid {\it Carl} 
  \; \not\rightarrow^* \;
  {\it SatiatedC} \mid P
\).
Note that the operator $\oplus$ will be exploited
in Sect.~\ref{sec:encodings}, when we will encode graph rewriting in
our calculus.

\end{example}

\section{Expressive power}\label{sec:encodings}

We now discuss the expressive power of our synchronization primitives, 
by showing how to encode some common concurrency idioms into our calculus.

\paragraph{Semaphores.}
Semaphores 
admit a simple
encoding in our calculus. 
Below, $n$ is the name associated with the
semaphore, while $x$ is a fresh variable. $P(n)$ and $V(n)$ denote
the standard semaphore operations, and process $Q$ is their continuation.
\begin{small}
\begin{align*}
& P(n).Q = (x) \; \fuse{x}{\mathsf{p}(n,x)}.Q \hspace{9mm}
V(n).Q = (x) \; \tell{\mathsf{p}(n,x)}.Q
\end{align*}
\end{small}
\noindent
Each $\fuse{x}{\mathsf{p}(n,x)}$ instantiates a variable $x$ such that
$\mathsf{p}(n,x)$ holds. Of course, the same $x$ cannot be
instantiated twice, so it is effectively consumed. 
New variables 
% to instantiate 
are furnished by $V(n)$.

\paragraph{Memory cells.}
We model below a memory cell. 
%The cell at any time contains a name $v$ as its value.
\begin{small}
\begin{align*}
\mathit{New}(n,v).Q&=(x)\tell{\mathsf{c}(n,x)\land \mathsf{d}(x,v)}.Q\\
\mathit{Get}(n,y).Q&=(w)\fuse{w}{\mathsf{c}(n,w)}.\join{y}{\mathsf{d}(w,y)}.\mathit{New}(n,y).Q \hspace{10mm}
\mathit{Set}(n,v).Q=(w)\fuse{x}{\mathsf{c}(n,w)}.\mathit{New}(n,v).Q
\end{align*}
\end{small}
The process $New(n,v)$ initializes the cell having name $n$ and initial
value $v$ (a name). 
The process $Get(n,y)$ recovers $v$ by fusing it with $y$: the
procedure is destructive, hence the cell is re-created. 
The process $Set(n,v)$ destroys the current cell and creates a new one.

\paragraph{Linda.}
Our calculus can model a tuple space, and implement the insertion
and retrieval of tuples as in Linda~\cite{Gelernter85linda}.
For illustration, we only consider
$\mathsf{p}$-tagged pairs here.
\begin{small}
\begin{align*}
\mathit{Out}(w,y).Q & =(x)\tell{\mathsf{p}(x)\land \mathsf{p}_{1}(x,w)\land \mathsf{p}_{2}(x,y)}.Q
&
\mathit{In}(w,y).Q  & = (x)\fuse{x}{\mathsf{p}_{1}(x,w)\land \mathsf{p}_{2}(x,y)}.Q\\
\mathit{In}(?w,y).Q &=(x)\fuse{x}{\mathsf{p}_{2}(x,y)}.\join{w}{\mathsf{p}_{1}(x,w)}.Q
&
\mathit{In}(w,?y).Q &=(x)\fuse{x}{\mathsf{p}_{1}(x,w)}.\join{y}{\mathsf{p}_{2}(x,y)}.Q\\
\mathit{In}(?w,?y).Q&=(x)\fuse{x}{\mathsf{p}(x)}.\join{w}{\mathsf{p}_{1}(x,w)}.\join{y}{\mathsf{p}_{2}(x,y)}.Q
\end{align*}
\end{small}
The operation $Out$ inserts a new pair in the tuple space. 
A fresh variable $x$ is related to the pair components through suitable
predicates. 
The operation $\mathit{In}$ retrieves a pair by pattern matching.
The pattern $\mathit{In}(w,y)$ mandates an exact match, so we require that both
components are as specified. 
Note that $\fuse{}{\!}$ will
instantiate the variable $x$, effectively consuming the tuple.  
The pattern $\mathit{In}(?w,y)$ requires to match only
against the $y$ component. 
We do exactly that in the $\fuse{}{\!}$ prefix. 
Then, we use $\join{}{\!}$ to recover the first component of the pair, 
and bind it to~$w$. 
The pattern $\mathit{In}(w,?y)$ is symmetric. 
The pattern $\mathit{In}(?w,?y)$
matches any pair, so we specify a weak requirement for the fusion.
Then we recover the pair components.

\paragraph{Synchronous $\pi$-calculus.}
We encode the synchronous $\pi$-calculus~\cite{Milner92pi} into our
calculus as follows:
\begin{small}
\begin{align*}
& [P | Q]  = [P] \; | \; [Q] \hspace{1cm}
[(\nu n) P] = (n) [P] \hspace{1cm}
[X(\vec{a})] = X(\vec{a}) \hspace{14mm}
[X(\vec{y}) \doteq P] = X(\vec{y}) \doteq [P] \\
&[\bar{a}\langle b\rangle.P]  = (x) \big( {\mathsf{msg}(x,b)} \; | \; \fuse{x}{\mathsf{in}(a,x)}.\; [P] \big) 
\hspace{13mm} \text{($x$ fresh)} \\
& [a(z).Q] =(y) \big( {\mathsf{in}(a,y)} \; | \; (z)\join{z}{\mathsf{msg}(y,z)}.\; [Q] \big) 
\hspace{10mm} \text{($y$ fresh)}
\end{align*}
\end{small}
Our encoding preserves parallel composition, and
maps name restriction to name delimitation, as one might desire.
The output cannot proceed with $P$ until $x$ is fused with some $y$.
Dually, the input cannot proceed until $y$ is instantiated to a name,
that is until $y$ is fused with some $x$ -- otherwise, there is no
way to satisfy $\mathsf{msg}(y,z)$.

The encoding above satisfies the requirements of
\cite{Gorla08encodability}.
It is {\em compositional}, mapping each $\pi$ construct in a context
of our calculus. 
Further, the encoding is {\em name invariant} and
preserves {\em termination, divergence} and {\em success}. Finally it
is {\em operationally corresponding} since, writing $\rightarrow_\pi$
for reduction in $\pi$,
\begin{align*}
 P \rightarrow_\pi^* P' &\implies [P] \rightarrowtail^* \sim [P'] \\
 [P] \rightarrowtail^* Q &\implies \exists P'.\; Q \rightarrowtail^*
\sim [P'] \land P\rightarrow_\pi^* P'
\end{align*}
where $\sim$ is $\rightarrowtail$-bisimilarity. 
For instance, note that:
\begin{align*}
& [ (\nu m) ( n\langle m \rangle . P | n(z).Q )] 
\;\; \rightarrowtail^* \;\;
(m)(o) \big( {\mathsf{msg}(o,m)} | [P] | {\mathsf{in}(n,o)} | 
[Q]\{\nicefrac{m}{y}\}
\big)  \sim (m) [P | Q\{\nicefrac{m}{y}\} ]
\end{align*}
since the name $o$ is fresh and the constraints ${\mathsf{msg}(o,m)},
{\mathsf{in}(n,o)}$ do not affect the behaviour of $P,Q$. To see this,
consider the inputs and outputs occurring in $P,Q$. Indeed, in the
encoding of inputs, the $\fuse{x}{\!}$ prefix will instantiate $x$ to
a fresh name, hence not with~$o$. On the other hand, in the encoding
of outputs, the $\join{z}{\!}$ prefix can fire only after $y$ has been
fused with $x$, hence instantiated with a fresh name. The presence of
${\mathsf{msg}(o,m)}$ has no impact on this firing.

Note that our encoding does not handle non-deterministic choice.
This is however manageable through 
the very same operator~$\oplus$ of Ex.~\ref{ex:oplus}. 
We will also exploit~$\oplus$ below, to encode graph rewriting.

\subsection{Graph rewriting} % \label{ex:graph-rew}
In the encoding of the $\pi$-calculus we have modelled a simple interaction
pattern; namely, Milner-style synchronization.  Our calculus is also
able to model more sophisticated synchronization mechanisms, such as
those employed in graph rewriting techniques~\cite{Rozenberg97handbook}.
Before dealing with the general case, we introduce our encoding
through a simple example. 

\begin{example}\label{ex:ring-rew}
Consider the following ``ring-to-star'' graph rewriting rule, inspired from an example
in~\cite{Lanese07mapping}:
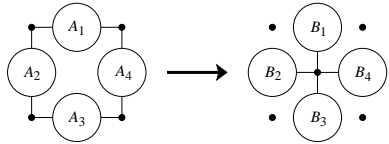
\begin{center}%
\begin{picture}(0,0)%
\includegraphics{graph.pstex}%
\end{picture}%
\setlength{\unitlength}{1657sp}%
\begingroup\makeatletter\ifx\SetFigFont\undefined%
\gdef\SetFigFont#1#2#3#4#5{%
  \reset@font\fontsize{#1}{#2pt}%
  \fontfamily{#3}\fontseries{#4}\fontshape{#5}%
  \selectfont}%
\fi\endgroup%
\begin{picture}(5686,2084)(983,-1778)
\put(2611,-796){\makebox(0,0)[lb]{\smash{{\SetFigFont{6}{7.2}{\rmdefault}{\mddefault}{\updefault}{\color[rgb]{0,0,0}$A_4$}%
}}}}
\put(1913,-121){\makebox(0,0)[lb]{\smash{{\SetFigFont{6}{7.2}{\rmdefault}{\mddefault}{\updefault}{\color[rgb]{0,0,0}$A_1$}%
}}}}
\put(1914,-1486){\makebox(0,0)[lb]{\smash{{\SetFigFont{6}{7.2}{\rmdefault}{\mddefault}{\updefault}{\color[rgb]{0,0,0}$A_3$}%
}}}}
\put(1246,-803){\makebox(0,0)[lb]{\smash{{\SetFigFont{6}{7.2}{\rmdefault}{\mddefault}{\updefault}{\color[rgb]{0,0,0}$A_2$}%
}}}}
\put(5528,-136){\makebox(0,0)[lb]{\smash{{\SetFigFont{6}{7.2}{\rmdefault}{\mddefault}{\updefault}{\color[rgb]{0,0,0}$B_1$}%
}}}}
\put(4846,-804){\makebox(0,0)[lb]{\smash{{\SetFigFont{6}{7.2}{\rmdefault}{\mddefault}{\updefault}{\color[rgb]{0,0,0}$B_2$}%
}}}}
\put(6196,-811){\makebox(0,0)[lb]{\smash{{\SetFigFont{6}{7.2}{\rmdefault}{\mddefault}{\updefault}{\color[rgb]{0,0,0}$B_4$}%
}}}}
\put(5529,-1486){\makebox(0,0)[lb]{\smash{{\SetFigFont{6}{7.2}{\rmdefault}{\mddefault}{\updefault}{\color[rgb]{0,0,0}$B_3$}%
}}}}
\end{picture}%
\end{center}
Whenever the processes $A_1 \ldots A_4$ are in a configuration matching 
the left side of the rule (where the bullets represent shared names)
a transition is enabled, leading to the right side. 
The processes change to $B_1 \ldots B_4$, and a fresh name is 
shared among all of them, while the old names are forgotten.
Modelling this kind of synchronization in, e.g., the $\pi$-calculus
would be cumbersome, since a discovery protocol must be devised to
allow processes to realize the transition is enabled. 
Note that no process is directly connected to the others, 
so this protocol is non-trivial.

Our calculus allows for an elegant, symmetric translation of the rule
above, which is interpreted as an agreement among the processes
$A_1\ldots A_4$. 
Intuitively, each process $A_i$ promises to change into $B_i$, 
and to adjust the names, provided that all the others perform 
the analogous action.
Since each $A_i$ shares two names with the other processes, 
we write it as $A_i(n,m)$. 
The advertised contract is specified below as a~\pcl formula, where we denote 
addition and subtraction modulo four as $\boxplus$ and $\boxminus$, respectively:
\begin{flalign}
& a_i(n,m,x) = \mathsf{f}_{i\boxplus 1}(x,m) \land \mathsf{s}_{i\boxminus 1}(x,n) 
 \coimp \mathsf{f}_i(x,n) \land \mathsf{s}_i(x,m)
\label{eq:contract-ai}
\end{flalign}
An intuitive interpretation of $\mathsf{f,s}$ is as follows:
$\mathsf{f}_i(x,n)$ states that $n$ is the {\em first} name of some
process $A_i(n,-)$ which is about to apply the rule. Similarly for
$\mathsf{s}_i(x,m)$ and the {\em second} name. The parameter $x$
is a session ID, uniquely identifying the current transition.
The contract $a_i(n,m,x)$ states that $A_i$ agrees to fire the rule
provided both its neighbours do as well. The actual $A_i$ process is as
follows.
\begin{align*}
A_i(n,m) &\doteq (x) \tell {a_i(n,m,x)} . 
 \fuse{x}{\mathsf{f}_i(x,n) \land \mathsf{s}_i(x,m) }
      . \, B_i(x)
\end{align*}
Our \pcl logic enables the wanted transition:
\(
P = |\! |_i \, A_i(n_i,n_{i\boxplus 1}) \rightarrowtail^*
(m) |\! |_i \, B_i(m)
\).

Note that the above works even when nodes $n_i$ are shared
among multiple parallel copies of the same processes. For instance,
$P|P$ will fire the rule twice, possibly mixing $A_i$ components
between the two $P$'s.
\end{example}

\paragraph{General Case. }
We now deal with the general case of a graph rewriting system.

\begin{definition}
An hypergraph $G$ is a pair $(V_G,E_G)$ where $V_G$ is a set of vertices
and $E_G$ is a set of hyperedges.
Each hyperedge $e\in E_{G}$ has an associated tag $tag(e)$ and 
an ordered tuple of vertices $(e_1,\ldots,e_k)$ where $e_j \in V_{G}$.
The tag $tag(e)$ uniquely determines the arity $k$.
\end{definition}

\begin{definition}\label{def:graph-rew-sys}
A graph rewriting system is a set of graph rewriting rules $\{ G_i
\Rightarrow H_i \}_i$ where $G_i,H_i$ are the {\em source} and 
{\em target} hypergraphs, respectively. 
No rule is allowed to discard vertices, i.e.~$V_{G_i} \subseteq V_{H_i}$.
Without loss of generality, we require that the sets of hyperedges 
$E_{G_i}$ are pairwise disjoint.
\end{definition}

In Def.~\ref{def:graph-rew} below, we recall how 
to apply a rewriting rule $G \Rightarrow H$ to a
given graph $J$.  The first step is to identify an embedding $\sigma$ of $G$
inside $J$. The embedding $\sigma$ roughly maps  $H\setminus G$ to 
a ``fresh extension'' of $J$ (i.e.~to the part of the graph that
is created by the rewriting). Finally, we
replace $\sigma(G)$ with $\sigma(H)$. 

\begin{definition}\label{def:graph-rew}
Let $\{ G_i \Rightarrow H_i \}_i$ be a graph rewriting system,
and let $J$ be a hypergraph. An {\em embedding} $\sigma$ of  $G_i$ 
in $J$ is a function such that:
{\bf (1)} $\sigma(v)\in V_J$ for each $v\in V_{G_i}$, and
$\sigma(v)\not\in V_J$ for each $v\in V_{H_i} \setminus V_{G_i}$ ;
{\bf (2)} $\sigma(e)\in E_J$ for each $e\in E_{G_i}$, and
$\sigma(e)\not\in E_J$ for each $e\in E_{H_i} \setminus E_{G_i}$ ;
{\bf (3)} $\sigma(v)=\sigma(v') \implies v=v'$ for each 
      $v,v'\in V_{H_i} \setminus V_{G_i}$ ;
{\bf (4)} $\sigma(e)=\sigma(e') \implies e=e'$ for each
      $e,e'\in E_{G_i} \cup E_{H_i}$ ;
{\bf (5)} $tag(e) = tag(\sigma(e))$ for each $e\in E_{G_i} \cup E_{H_i}$ ;
{\bf (6)} $\sigma(e)_h = \sigma(e_h)$ for each $e\in E_{G_i} \cup E_{H_i}$
and $1 \leq h \leq k$.

%An extended embedding $\hat{\sigma}$ is an extension of $\sigma$ to
%$H_i$ such that 
%$\hat{\sigma}(V_{H_i} \setminus V_{G_i}) \cap V_K = \emptyset$
%and $\hat{\sigma}(E_{H_i} \setminus E_{G_i}) \cap E_K = \emptyset$.

The {\em rewriting relation} $J\rew K$ holds iff, for some embedding
$\sigma$, we have 
$V_K = (V_J \setminus \sigma(V_{G_i})) \cup \sigma(V_{H_i})$ and
$E_K = (E_J \setminus \sigma(E_{G_i})) \cup \sigma(E_{H_i})$.
The assumption 
$V_{G_i} \subseteq V_{H_i}$ of Def.~\ref{def:graph-rew-sys} ensures
$V_J \subseteq V_K$, so no dangling hyperedges are created
by rewriting.
\end{definition}

We now proceed to encode graph rewriting in our calculus.
To simplify our encoding, we make a mild assumption: we require each
$G_i$ to be a {\em connected} hypergraph. Then, encoding a generic
hypergraph $J$ is performed in a compositional way: 
we assign a unique name $n$ to each vertex in $V_J$,
and then build a parallel composition of processes
$A_{tag(e)}(\vec{n})$, one for each hyperedge $e$ in $E_J$, where
$\vec{n}=(n_1,\ldots,n_k)$ identifies the adjacent vertices. Note
that since the behaviour of an hyperedge $e$ depends on its tag, only,
we index $A$ with $t=tag(e)$.
Note that $t$ might be the tag of several
hyperedges in each source hypergraph $G_i$. We stress this point: tag $t$
may occur in distinct source graphs $G_i$, and each of these may have
multiple hyperedges tagged with $t$.  The process $A_t$ must then be able
to play the role of any of these hyperedges. The willingness to play
the role of such a hyperedge $e$ relatively to a single node $n$ is
modelled by a formula $\mathsf{p}_{e,h}(x,n)$ meaning ``I agree to
play the role of $e$ in session $x$, and my $h$-th node is $n$''.  
The session variable $x$ is exploited to ``group'' all the constraints
related to the same rewriting. We use the formula $\mathsf{p}_{e,h}(x,n)$
in the definition of $A_t$. The process $A_t(\vec{n})$ promises
$\mathsf{p}_{e,1}(x,n_1), \ldots, \mathsf{p}_{e,k}(x,n_k)$
(roughly, ``I agree to be rewritten as $e$''), provided that
all the other hyperedges sharing a node $n_h$
agree to be rewritten according to their roles $\bar{e}$. Formally, the 
contract related to $e\in E_{G_i}$ is the following:
\begin{flalign}
& a_e(x,% w,
\vec{n}) \hspace{20pt} =
\bigwedge_{
   \scriptsize
   \begin{array}{c}
   1\leq h\leq k, \;
   \bar{e}\in E_{G_i}, \;
   \bar{e}_{\bar{h}} = e_h
   \end{array}
       } 
     \mathsf{p}_{\bar{e},\bar{h}}(x,n_h) \coimp 
% \mathsf{go}_e(w) \land
\bigwedge_{
  % \footnotesize
  % \begin{array}{c}
  % h\, s.t. \\
   1\leq h\leq k
  %\end{array}
} \mathsf{p}_{e,h}(x,n_h)
\label{eq:contract-ae}
\end{flalign}

Note that in the previous example we indeed followed this schema of
contracts. There, the hypergraph $J$ has four hyperedges $e_1$, $e_2$,
$e_3$, $e_4$, each with a unique tag.
The formulae $\mathsf{f}_i$ and $\mathsf{s}_i$ in \eqref{eq:contract-ai}
are rendered as $\mathsf{p}_{e_i,1}$ and $\mathsf{p}_{e_i,2}$ in
\eqref{eq:contract-ae}. Also the operators $\boxplus 1$ and $\boxminus 1$,
used in \eqref{eq:contract-ai} to choose neighbours,
are generalized in \eqref{eq:contract-ae} 
through the condition $\bar{e}_{\bar{h}} = e_h$.

Back to the general case, the process $A_t$ will advertise the 
contract $a_e$ for each $e$ having tag $t$, and then will 
try to fuse variable $x$.
Note that, since the neighbours are advertising the analogous
contract, we can not derive any $\mathsf{p}_{e,h}(x,n_h)$ unless
{\em all} the hyperedges in the connected component agree to be rewritten.
Since $G_i$ is connected by hypothesis, this means that we indeed
require the whole graph to agree.

However, advertising the contracts $a_e$ using a simple parallel
composition can lead to unwanted results when non-determinism is
involved. Consider two unary hyperedges, which share a node $n$,
and can be rewritten using two distinct rules:
$G \Rightarrow H$ with $e1,e2 \in E_G$, and
$\bar{G} \Rightarrow \bar{H}$ with $\bar e1, \bar e2 \in E_{\bar{G}}$. 
Let $tag(e1)=tag(\bar e1)=t1$ and $tag(\bar e2)=tag(\bar e2)=t2$.
Each process thus advertises two contracts, \text{e.g.}:
\begin{align*}
& A_{t1}  = (x) \; (a_{e1}(x,n) \; | \; a_{\bar e1}(x,n) \; | \; Fusion_{t1} ) 
&& A_{t2} = (x) \; (a_{e2}(x,n) \; | \; a_{\bar e2}(x,n) \; | \; Fusion_{t2} ) 
\end{align*}

Consider now $A_{t1} \; | \; A_{t2}$. 
After the fusion of $x$, it is crucial that both hyperedges agree on
the rewriting rule that is being applied -- that is
either they play the roles of $e1,e2$ or those of $\bar e1, \bar e2$.
However, only one
$Fusion$ process above will perform the fusion, say e.g.\ the first one
(the name $m$ below is fresh):
\[
(m) (a_{e1}(m,n) | a_{\bar e1}(m,n) | Rewrite_{e1} |
     a_{e2}(m,n) | a_{\bar e2}(m,n) | Fusion_{t2}\{\nicefrac{m}{x}\} ) 
\]
Note that the process 
$Fusion_{t2}\{\nicefrac{m}{x}\}$ can still
proceed with the {\em other} rewriting, since the substitution above
cannot disable a prefix which was enabled before. 
So, we can end up with $Rewrite_{\bar e2}$, leading to an inconsistent
rewriting. Indeed, $A_{t1}$ was rewritten using $G \Rightarrow H$,
while $A_{t2}$ according to $\bar G \Rightarrow \bar H$.

To avoid this, we resort to the construction $\oplus_i p_i$
discussed in Ex.~\ref{ex:oplus}. We can then define $A_t$ as follows.
\[
A_t(\vec{n}) \doteq
  (x) ( \bigoplus_{
       % \begin{array}{c}
       tag(e) = t
       % \end{array}
   } a_e(x,\vec{n}) | \sum_{tag(e)=t} \fuse{x}{
\bigwedge_{1\leq h\leq k} \mathsf{p}_{e,n}(x,n_h)
%\mathsf{go}_e(w)
}.B_e(x,\vec{n}) )
\]
In each $A_t$, the contracts $a_e$ are exposed under the $\oplus$.
The consequences of these contracts are then demanded by a sum of
$\fuse{x}{}$. We defer the definition of $B_e$.

Consider now the behaviour of the encoding of a whole hypergraph:
$A_t(\vec{n}) | \cdots | A_{t'}(\vec{n}')$.  If the hypergraph $J$
contains an occurrence of $G$, where $G \Rightarrow H$ is a rewriting
rule, each of the processes involved in the occurrence
$P_1,\ldots,P_l$ may fire a $\fuse{x}{}$ prefix. Note that this prefix
demands {\em exactly one} contract $a_e$ from each process inside of
the occurrence of $G$. This is because, by construction, each $a_e$
under the same $\oplus$ involves distinct $\mathsf{p}_{e,n}$.
This implies that, whenever a fusion is performed, the contracts
which are not directly involved in the handshaking, but are present
in the occurrence of $G$ triggering the rewriting, are then effectively
disabled. In other words, after a fusion the sums in the other involved
processes have exactly one enabled branch, and so they are now
committed to apply the rewriting coherently.

After the fusion $B_e(x,\vec{n})$ is reached, where $x$ has been
instantiated with a fresh session name $m$ which is common to all the
participants to the rewriting. It is then easy to exploit this name $m$ 
to reconfigure the graph. Each involved vertex (say, with name $n$)
can be exposed to all the participants through
e.g.~$\tell{\mathsf{vert}_h(m,n)}$, and retrieved through the
corresponding $\join{y}{\mathsf{vert}_h(m,y)}$. Since $m$ is fresh,
there is no risk of interference between parallel rewritings.  New
vertices (those in $V_H\setminus V_G$) can be spawned and broadcast in
a similar fashion. Once all the names are shared, the target
hypergraph $H$ is formed by spawning its hyperedges $E_H$ through a
simple parallel composition of $A_t(\vec{n})$ processes -- each one
with the relevant names. Note that the processes
$A_t$, where $t$ ranges over all the tags, are mutually recursive.

{\bf Correctness.}  Whenever we have a rewriting $J\rightarrow K$, it
is simple to check that the contracts used in the encoding yield an
handshaking, so causing the corresponding transitions in our process
calculus. The reader might wonder whether the opposite also holds,
hence establishing an {\em operational correspondence}. It turns out
that our encoding actually allows {\em more} rewritings to take place, with
respect to Def.~\ref{def:graph-rew}. Using the $A_i$ from 
Ex.~\ref{ex:ring-rew}, we have that the following loop of length 8
can perform a transition.
\[
P= A_1(n_1,n_2) |
A_2(n_2,n_3) |
A_3(n_3,n_4) |
A_4(n_4,n_5) |
A_1(n_5,n_6) |
A_2(n_6,n_7) |
A_3(n_7,n_8) |
A_4(n_8,n_1) 
\]
Indeed, any edge here has exactly the same ``local view'' of the graph
as the corresponding $G$ of the rewriting rule. So, an handshaking takes
place. Roughly, if a graph $J_0$ triggers a rewriting in the encoding,
then each ``bisimilar'' graph $J_1$ will trigger the same rewriting.

A possible solution to capture graph rewriting in an exact way would
be to mention all the vertices in each contract. That is, edge $A_1$
would use $p_{A_1}(n_1,n_2,x,y)$, while edge $A_2$ would use
$p_{A_2}(w,n_2,n_3,z)$, and so on, using fresh variables for each
locally-unknown node. Then, we would need the $\fuse{}{}$ prefix to
match these variables as well, hence precisely establishing the
embedding $\sigma$ of Def.~\ref{def:graph-rew}. 
The semantics of $\fuse{}{}$ introduced in ~\cite{BZ10lics}
allows for such treatment.

%%% Local Variables: 
%%% mode: latex
%%% TeX-master: "main"
%%% End: 

\section{Discussion} \label{sec:related-work}

We have investigated primitives for contract-based interaction.
Such primitives extend those of Concurrent Constraints, by allowing
a multi-party mechanism for the fusion of variables
which well suites to model contract agreements.
We have shown our calculus expressive enough to model a variety of 
typical contracting scenarios.
To do that, we have also exploited our propositional contract logic 
\pcl~\cite{BZ10lics} to deduce the duties inferred
from a set of contracts.
Finally, we have encoded into our calculus some common idioms for concurrency,
among which the $\pi$-calculus and graph rewriting.

Compared to the calculus in~\cite{BZ10lics}, the current one features
a different rule for managing the fusion of variables.
In~\cite{BZ10lics}, the prefix $\fuse xc$ picks from the context a set
of variables $\vec{y}$ (like ours) and a set of {\em names} $\vec{m}$
(unlike ours). Then, the (minimal) fusion $\sigma$ causes the
variables in $x\vec{y}$ to be replaced with names in $n\vec{m}$, where
$n$ is fresh, and $\sigma(x)=n$. 
The motivation underneath this complex fusion mechanism is that, 
to establish a session in~\cite{BZ10lics}, 
we need to instantiate the variables $\vec{y}$
which represent the identities of the  principals involved in the handshaking.
% before we check for entailment.
Similarly, $\join{\vec{x}}{c}$ is allowed to instantiate a set of
variables $\vec{x}$. Instead, in this paper, to present our
expressivity results we have chosen a simplified version of the
calculus, where $\fuse xc$ considers a single name, and $\join xc$ a
single variable. 
At the time of writing, we do not know whether the simplified calculus
presented here is as expressive as the calculus of~\cite{BZ10lics}. 
The contract-related examples shown in this paper
did not require the more sophisticated rules for $\fuse{}{}$, 
nor did the encodings of most of the concurrency idioms. 
 As a main exception, we were unable to perfectly encode graph rewriting 
in our simplified calculus; the difficulty there was that of distinguishing 
between bisimilar graphs.
We conjecture that the more general fusion of~\cite{BZ10lics} 
is needed to make the encoding perfect; proving this  
would show our simplified calculus strictly less expressive than the one
in~\cite{BZ10lics}.

In our model of contracts we have abstracted from most 
of the implementation issues.
For instance, in insecure environments populated by attackers,
the operation of exchanging contracts requires particular care.
Clearly, integrity of contracts is a main concern, so we expect that
suitable mechanisms have to be applied to ensure that contracts are not 
tampered with. 
Further, establishing an agreement between participants in a
distributed system with unreliable communications appears similar
to establishing \emph{common knowledge} among the 
stipulating parties~\cite{Halpern90jacm}, so an implementation
has to cope with the related issues. For instance, the 
$\fuse{x}{}$ prefix requires a fresh name to be delivered 
among all the contracting parties, so the implementation must ensure
everyone agrees on that name. Also, it is important that
participants can be coerced to respect their contracts after the
stipulation: to this aim, the implementation should at least ensure
the non repudiation of contracts~\cite{Zhou01nonrepudiation}.

Negotiation and service-level agreement are dealt with in 
cc-pi~\cite{Buscemi07ccpi,Buscemi08ccpi}, 
a calculus combining features from concurrent constraints and name passing;
\cite{Buscemi07transactional} adds rules for handling transactions.
As in the $\pi$-calculus, synchronization is channel-based:
it only happens between two processes sharing a name.
Synchronization fuses two names, similarly to the 
fusion calculus and ours.
A main difference between cc-pi and our calculus
is that in cc-pi only two parties may simultaneously reach an agreement,
while our $\fuse{}{\!}$ allows for simultaneous multi-party agreements.
Also, in our calculus the parties involved in an agreement do not 
have to share a pre-agreed name.
This is useful for modelling scenarios where a contract can be accepted
by any party meeting the required terms (see e.g.\ Ex.~\ref{ex:proc:judge}).

In~\cite{Castagna09contracts} contracts are CCS-like processes.
A client contract is compliant with a service contract if 
any possible interaction between the client and the service will always succeed,
i.e.\ all the expected synchronizations will take place.
% There, a main problem is how to define (and decide) a subcontract relation,
% that allows for safely substituting services without affecting the
% compliance with their clients.
This is rather different from what we expect from a calculus for contracts.
For instance, consider a simple buyer-seller scenario.
In our vision, it is important to provide the buyer with the guarantee
that, after the payment has been made, then either the payed goods
are made available, or a refund is issued. 
Also, we want to assure the seller that
a buyer will not repudiate a completed transaction.
We can model this by the following contracts in \pcl\!:
$\mathit{Buyer} = (\mathsf{ship} \lor \mathsf{refund}) \coimp \mathsf{pay}$,
and
$\mathit{Seller} = \mathsf{pay} \coimp (\mathsf{ship} \lor \mathsf{refund})$.
Such contracts lead to an agreement.
The contracts of~\cite{Castagna09contracts}
would have a rather different form, \text{e.g.}
${\it Buyer} = \overline{\mathsf{pay}}.\, ( \mathsf{ship} + \mathsf{refund} )$
and
${\it Seller} = \mathsf{pay}.\, (\overline{\mathsf{ship}} \oplus \overline{\mathsf{refund}})$,
where~$+$ and~$\oplus$ stand respectively for external and internal choice.
This models the client outputting a payment, and then
either receiving the item or a refund (at service discretion).
Dually, the service will first input a payment, and then opt for shipping the
item or issuing a refund.
This is quite distant from our notion of contracts.
Our contracts could be seen as a declarative underspecified description 
of which behavioural contracts are an implementation.
Behavioural contracts seem more rigid than ours, as they 
precisely fix the order in which the actions must be performed.
Even though in some cases this may be desirable, many real-world contracts
allow for a more liberal way of constraining the involved parties
(\text{e.g.}, ``I will pay before the deadline'').
While the crucial notion in~\cite{Castagna09contracts} 
is \emph{compatibility} (which results in a yes/no output), 
we focus on the inferring the \emph{obligations} that arise from 
a set of contracts. 
This provides a fine-grained quantification of the reached agreement,
e.g.\ we may identify who is responsible of a contract violation.

Our calculus could be exploited to enhance the compensation mechanism
of long-running transactions~\cite{Bocchi03calculus,Bruni05theoretical,Butler04trace}.
There, a transaction is partitioned  into a sequence of
smaller ones, each one associated with a
\emph{compensation}, to be run upon failures of the standard execution.
While in long-running transactions clients have little control 
on the compensations (specified by the designer),
in our approach clients can use contracts to select those services
offering the desired compensation.

An interesting line for future work is that of comparing the
expressiveness of our calculus against other general synchronization
models. Our synchronization mechanism, based on the local minimal
fusion policy and PCL contracts, seems to share some similarities with
the synchronization algebras with mobility~\cite{Lanese05shr}.
Indeed, in many cases, it seems to be possible to achieve the
synchronization defined by a SAM through some handshaking in our
model. We expect that a number of SAMs can be encoded through suitable
PCL contracts, without changing the entailment relation.
Dually, we expect that the interactions deriving from a set of
contracts could often be specified through a SAM. 

Another general model for synchronization is the BIP
model~\cite{Bliudze08bip}. Here, complex coordination schemes can be
defined through an algebra of connectors. While some of these schemes
could be modelled by contracts, encoding BIP priorities into our
framework seems to be hard.  Actually, the only apparent link between
priorities and our calculus is the minimality requirement on fusions.
However, our mechanism appears to be less general. For instance,
BIP allows maximal progress as its priority relation, which contrasts
with the minimality of our fusions.

\paragraph{Acknowledgments.}
Work partially supported by
MIUR Project {\sc SOFT}, and
Autonomous Region of Sardinia Project {\sc TESLA}.

\bibliographystyle{plain}
\bibliography{logic}

\end{document}